\documentclass[aps,prd,preprint,showpacs,floatfix,nofootinbib]{revtex4-1}

\usepackage{dcolumn}
\usepackage{amsmath, amsfonts, amssymb, mathrsfs, times, float, color, bm}
\usepackage{extarrows}
\usepackage{graphicx}
\usepackage{graphicx,epstopdf}
\usepackage{dcolumn}
\usepackage{bm}
\newcommand{\nn}{\nonumber}

\begin{document}

\title{Exact Harmonic Metric for a Moving Reissner-Nordstr\"{o}m Black Hole}
\author{G. He and W. Lin
}\thanks{To whom correspondence should be addressed. Email: wl@swjtu.edu.cn}

\affiliation{
School of Physical Science and Technology, Southwest Jiaotong University, Chengdu 610031, China}
\begin{abstract}
The exact harmonic metric for a moving Reissner-Nordstr\"{o}m black hole with an arbitrary constant speed is presented. As an application, the post-Newtonian dynamics of a non-relativistic particle in this field is calculated.

\end{abstract}

\maketitle

\section{Introduction}
The motion of a gravitational source can affect the dynamics of particle passing by it, and this effect has attracted considerable attention over the last two decades
~\cite{LTK1974,PyneBirk1993,KopeiSch1999,Sereno2002b,MQA2003,Sereno2005b,Sereno2007,KopeiMak2007,Bonvin2008,Novatietal2010,ZscKlio2011,KopeiMash2002}. There are several methods to calculate this effect. One is to directly solve Li\'{e}nard-Wiechert gravitational potential from the field equations, which has been used to study light propagation in the gravitational field of an arbitrarily moving N-body system, as well as that with angular momentum~\cite{KopeiSch1999,KopeiMash2002}. Another method takes advantage of the general covariance of field equations to obtain the metric of the moving source from the known static source's metric via Lorentz transformation~\cite{WuckSper2004}. Recently, this method was employed to derive the time-dependent harmonic metrics of arbitrary-constant moving Schwarzschild and Kerr black holes~\cite{LH2014,LH2014b,LHJ2014}.

In this work, we apply a Lorentz transformation to derive the exact harmonic metric for a moving Reissner-Nordstr\"{o}m black hole with an arbitrary constant speed. Furthermore, based on the metric, we calculate the dynamics of a photon and a particle in the weak-field limit. In what follows we use geometrized units $(G=c=1)$.

\section{Exact Harmonic Metric for an Arbitrarily Constantly Moving Reissner-Nordstr\"{o}m Black Hole}\label{secion_HarmonicRN}
We start with the harmonic metric of Reissner-Nordstr\"{o}m black hole, which can be written as~\cite{LJ2014}
\begin{eqnarray}
ds^{2}=-\frac{R^2-m^2+Q^2}{(R+m)^2}dX_0^2+\left(1+\frac{m}{R}\right)^2
\left[\delta_{ij}+\frac{m^2-Q^2}{R^2-m^2+Q^2}\frac{X_iX_j}{R^2}\right]dX_idX_j~,~~\label{HarmonicRN}
\end{eqnarray}
where $m$ and $Q$ are the rest mass and electric charge of the black hole, respectively. $i,~j=1,~2,~3$, and $\delta_{ij}$ denotes Kronecker delta. Notice that here $X_{\mu}$ denotes the contravariant vector $x'^{\mu}=(t',~x',~y',~z')$ for display convenience, and $R^2=X_1^2+X_2^2+X_3^2$.

Since Einstein field equations have the property of general covariance, the harmonic metric of a constantly moving R-N black hole can be obtained via applying a Lorentz boost to Eq. \eqref{HarmonicRN}. We denote the coordinate frame of the background as $(t,~x,~y,~z)\,$, and assume the velocity of the black hole to be ${\bm v}=v_1\bm{e}_1+v_2\bm{e}_2+v_3\bm{e}_3~$, with $\bm{e}_i~(i=1,2,3)$ denoting the unit vector of 3-dimensional Cartesian coordinates. The Lorentz transformation between $(t,~x,~y,~z)$ and the comoving frame $(t',~x',~y',~z')$ of the moving hole can be written as
\begin{equation}
x'^\alpha = \Lambda^\alpha_\beta x^\beta~, \label{LorentzTran0}
\end{equation}
with
\begin{eqnarray}
&&\Lambda_0^0=\gamma~,  \label{LorentzTran1} \\
&&\Lambda_0^i=\Lambda_i^0=-v_i\gamma~,  \label{LorentzTran2} \\
&&\Lambda^i_j=\delta_{ij}+v_iv_j\frac{\gamma-1}{v^2}~,  \label{LorentzTran3}
\end{eqnarray}
where $\gamma= (1-v^2)^{-\scriptstyle \frac{1}{2}}$ is the Lorentz factor and $v^2=v_1^2+v_2^2+v_3^2$. Therefore, the exact harmonic metric of the moving Reissner-Nordstr\"{o}m black hole can be obtained as follows
\begin{eqnarray}
&&g_{00}=-\frac{\gamma^2(R^2-m^2+Q^2)}{(R+m)^2}
+\gamma^2\left(1+\frac{m}{R}\right)^2\left[\bm{v}^2+\frac{(\bm{v}\cdot\bm{X})^2(m^2-Q^2)}{R^2(R^2-m^2+Q^2)}\right]~,  \label{g00mRN}   \\
\nn&&g_{0i}=v_i\gamma^2\left[\frac{R^2\!-\!m^2\!+\!Q^2}{(R+m)^2}\!-\!\left(\!1\!+\!\frac{m}{R}\right)^2\right]
\!-\!\gamma\!\left(\!1\!+\!\frac{m}{R}\right)^2\!\!\frac{m^2-Q^2}{R^2(R^2\!-\!m^2\!+\!Q^2)}\times  \\
&&\hspace*{22pt}\left[\!X_i(\bm{v}\!\cdot\!\!\bm{X})\!+\!\frac{v_i(\gamma\!-\!1)(\bm{v}\!\cdot\!\!\bm{X})^2}{\bm{v}^2}\right]~,  \label{g0imRN}    \\
\nn&&g_{ij}=\left(1+\frac{m}{R}\right)^2\!\left\{\delta_{ij}+\frac{m^2-Q^2}{R^2(R^2-m^2+Q^2)}
\left[X_i+\frac{v_i(\gamma-1)(\bm{v}\cdot\bm{X})}{\bm{v}^2}\right]\times \right. \\
&&\hspace*{22pt}\left.\left[X_j\!+\!\frac{v_j(\gamma-1)(\bm{v}\cdot\bm{X})}{\bm{v}^2}\right]\!\right\}
\!+\!v_iv_j\gamma^2\left[\left(1+\frac{m}{R}\right)^2-\frac{R^2-m^2+Q^2}{(R+m)^2}\right]~. \label{gijmRN}
\end{eqnarray}
If we set $Q=0$, Eqs. \eqref{g00mRN} - \eqref{gijmRN} reduce to the harmonic metric of a moving Schwarzschild black hole with velocity $\bm v$
\begin{eqnarray}
&&g_{00}=-\frac{\gamma^2(1+\Phi)}{1-\Phi}+\bm{v}^2\gamma^2(1-\Phi)^2+\frac{\gamma^2\Phi^2(1-\Phi)}{1+\Phi}\frac{(\bm{v}\cdot\bm{X})^2}{R^2}~,\label{g00mS}\\
&&g_{0i}=v_i\gamma^2\!\left[\frac{1+\Phi}{1-\Phi}\!-\!(1\!-\!\Phi)^2\right]\!-\!\frac{\gamma\Phi^2(1-\Phi)}{1+\Phi}
\left[\frac{X_i(\bm{v}\cdot\bm{X})}{R^2}+\frac{v_i(\gamma-1)(\bm{v}\cdot\bm{X})^2}{\bm{v}^2R^2}\right]~,~~~~~~\label{g0imS}    \\
\nn&&g_{ij}=(1-\Phi)^2\,\delta_{ij}\!+\!\frac{\Phi^2(1-\Phi)}{R^2(1+\Phi)}\left[X_i\!+\!\frac{v_i(\gamma\!-\!1)(\bm{v}\!\cdot\!\bm{X})}{\bm{v}^2}\right]\!
\left[X_j\!+\!\frac{v_j(\gamma-1)(\bm{v}\!\cdot\!\bm{X})}{\bm{v}^2}\right] \\
&&\hspace*{25pt}+v_iv_j\gamma^2\!\left[(1-\Phi)^2\!-\!\frac{1+\Phi}{1-\Phi}\right]~,~~~~~~\label{gijmS}
\end{eqnarray}
which are the extension of the exact metric \cite{LH2014} for a Schwarzschild black hole with ${\bm v}=v {\bm e}_1$. Here $R$ is also equal to $\sqrt{X_1^2+X_2^2+X_3^2}$. It is worth pointing out that Eqs. \eqref{g00mS} - \eqref{gijmS}, to the first post-Minkowskian approximation, are in agreement with the gravitational Li\'{e}nard-Wiechert retarded solution~\cite{KopeiSch1999}.

\section{Dynamics of particle in the weak-field limit}\label{secion_ParticleDynamics}
As an application, we apply the harmonic metric to derive the post-Newtonian dynamics of a neutral and non-relativistic particle in the far field of the moving Reissner-Nordstr\"{o}m black hole. First, we expand Eqs. \eqref{g00mRN} - \eqref{gijmRN} up to an order of $1/R^2$
\begin{eqnarray}
&&g_{00}=-1-2(1+2v^2)\Phi-2\Phi^2-\frac{Q^2}{R^2}~,  \\
&&g_{0i}=4v_i \Phi~,  \\
&&g_{ij}=(1-2\Phi)\delta_{ij}~,
\end{eqnarray}
where the velocity of the black hole has also been assumed to be non-relativistic, i.e., $\gamma\simeq 1$. After tedious but straightforward calculations, up to the order of $\overline{v}^4/\overline{r}$ ($\overline{v}$ and $\overline{r}$ denote typical values of velocity and separation of a system of particles, respectively), we can obtain the equation of motion of a massive particle as follow
\begin{eqnarray}
\frac{d\bm{u}}{dt}=&&-\nabla\!\left(\!\Phi\!+\!2v^2\Phi\!+\!2\Phi^2\!+\!\frac{Q^2}{2R^2}\!\right)\!-\!\frac{\partial \bm{\zeta}}{\partial t}
\!+\!\bm{u}\!\times\!\left(\nabla \!\times\! \bm{\zeta}\right)\!+\!~\! 3 \bm{u}\frac{\partial \Phi}{\partial t} +  4 \bm{u}\left(\bm{u} \cdot \nabla\right)\Phi - \bm{u}^2\nabla\Phi~,~~~~~~~~\label{d5}
\end{eqnarray}
where ${\bm u}$ denotes the velocity of the particle, and $\bm{\zeta}=4\bm{v}\Phi$. When the charge of the black hole vanishes, this equation reduces to the post-Newtonian dynamics of a non-relativistic particle in the field of a moving Schwarzschild black hole~\cite{LH2014,Weinberg1972}.

\section{Conclusion}\label{secConclusion}
The metric in harmonic coordinates plays an important role in the post-Newtonian dynamics and gravitational wave radiation. In this work we obtain the exact metric for a moving Reissner-Nordstr\"{o}m black hole via applying a Lorentz boost to the Reissner-Nordstr\"{o}m metric in the harmonic coordinates. This method can avoid directly solving the Einstein field equations for a moving gravitational source. Based on this metric, we derive the post-Newtonian dynamics of a non-relativistic particle. This metric can also be used to calculate the deflection and time delay of light passing by a non-static Reissner-Nordstr\"{o}m black hole, as well as Hawking radiation of the black hole.

\section*{Acknowledgments}

This work was supported in part by the Program for New Century Excellent Talents in University (Grant No. NCET-10-0702), the National Basic Research Program of China (973 Program) (Grant No. 2013CB328904), and the Ph.D. Foundation of the Ministry of Education of China (Grant No. 20110184110016).


\begin{thebibliography}{60}

  \bibitem{LTK1974} K. H. Look, C. L. Tsou and H. Y. Kuo, {\it Acta Phys. Sin.} {\bf 23}, 225 (1974).
  \bibitem{PyneBirk1993} T. Pyne and M. Birkinshaw, {\it Astrophys. J.} {\bf 415}, 459 (1993).
  \bibitem{KopeiSch1999} S. M. Kopeikin and G. Sch\"{a}fer, {\it Phys. Rev. D} {\bf 60}, 124002 (1999).
  \bibitem{Sereno2002b} M. Sereno, {\it Phys. Lett. A} {\bf 305}, 7 (2002).
  \bibitem{KopeiMash2002} S. M. Kopeikin and B. Mashhoon, {\it Phys. Rev. D} {\bf 65}, 064025 (2002).
  \bibitem{MQA2003} M. Q. Miao, S. J. Qing and L. C. An, {\it Acta Phys. Sin.} {\bf 7}, 049 (2003).
  \bibitem{Sereno2005b} M. Sereno, {\it Mon. Not. R. Astron. Soc.} {\bf 359}, L19 (2005).
  \bibitem{Sereno2007} M. Sereno, {\it Mon. Not. R. Astron. Soc.} {\bf 380}, 1023 (2007).
  \bibitem{KopeiMak2007} S. M. Kopeikin and V. V. Makarov, {\it Phys. Rev. D} {\bf 75}, 062002 (2007).
  \bibitem{Bonvin2008} C. Bonvin, {\it Phys. Rev. D} {\bf 78}, 123530 (2008).
  \bibitem{Novatietal2010} S. C. Novati, M. Dall'Ora et al., {\it Astrophys. J.} {\bf 717}, 987 (2010).
  \bibitem{ZscKlio2011} S. Zschocke and S. A. Klioner, {\it Class. Quantum Grav.} {\bf 28}, 015009 (2011).
  \bibitem{WuckSper2004} O. Wucknitz and U. Sperhake, {\it Phys. Rev. D} {\bf 69}, 063001 (2004).


  \bibitem{LH2014} G. He and W. Lin, {\it Commun. Theor. Phys.} {\bf 61}, 270 (2014).
  \bibitem{LH2014b} G. He and W. Lin, {\it Int. J. Mod. Phys. D} {\bf 23}, 1450031 (2014).
  \bibitem{LHJ2014} G. He, C. Jiang and W. Lin, under review (2014).
  \bibitem{LJ2014} W. Lin and C. Jiang, {\it Phys. Rev. D} {\bf 89}, 087502 (2014).
  \bibitem{Weinberg1972} S. Weinberg, {\it Gravitation and Cosmology: Principles and Applications of the General Theory of Relativity}, Wiley, New York (1972).

\end{thebibliography}
\end{document}